\documentclass[a4paper,11pt]{article}
\usepackage{jinstpub} 
\usepackage{graphicx}
\usepackage{dcolumn}
\usepackage{bm}
\usepackage{subfigure,dcolumn}
\usepackage{hyperref}
\usepackage{pgfplots}
\usepackage{booktabs}
\usepackage{multirow}
\pgfplotsset{compat=1.18}

\title{Real-Time-Capable Betatron Tune Measurement from Schottky Spectra Using Deep Learning and Uncertainty-Aware Kalman Filtering}

\author[a,b]{P. Sun}
\author[a,c,1]{M. Zhang\note{Corresponding author.}}
\author[a,c]{R. Yuan}
\author[a]{D. Li}
\author[c]{J. Dong}
\author[d]{and Y. Shi}

\affiliation[a]{Shanghai Institute of Applied Physics, Chinese Academy of Sciences,\\
Shanghai 201800, China}
\affiliation[b]{University of Chinese Academy of Sciences,\\
China}
\affiliation[c]{Shanghai Advanced Research Institute, Chinese Academy of Sciences,\\
Shanghai 201204, China}
\affiliation[d]{Northwest Institute of Nuclear Technology (NINT), \\
Xi'an 710024, China}

\emailAdd{zhangmanzhou@sinap.ac.cn}

\abstract{
Betatron tune measurement is essential for beam control in compact proton-therapy synchrotrons, yet conventional peak-detection techniques are not robust under the low signal-to-noise ratio (SNR) conditions typical of these machines. This work presents a lightweight convolutional neural network that performs real-time tune extraction from Schottky spectra with sub-millisecond inference latency and calibrated uncertainty estimates. The model uses attention-based pooling for reliable peak localization and a dual-branch architecture that jointly predicts the tune and its associated uncertainty. Trained with a Laplace negative log-likelihood loss, it produces uncertainty estimates whose magnitude tracks the instantaneous prediction error, which enables uncertainty-aware Kalman filtering for temporal smoothing. Experiments on a large synthetic dataset spanning SNR levels from 0 to $-20$\,dB demonstrate substantial performance gains over traditional peak-detection baselines, while the Kalman filter further suppresses transient outliers in time-series operation. Preliminary validation on operational beam data confirms stable tune tracking without retraining. With only about $2.0\times 10^{4}$ trainable parameters and real-time inference on commodity GPU hardware, the proposed diagnostic offers a practical solution for rapid and accurate betatron tune monitoring in compact medical synchrotrons and similar accelerators.
}


\arxivnumber{2412.19171} 

\begin{document}
\maketitle
\flushbottom


\section{Introduction}
\label{sec:intro}

Betatron tune measurement is fundamental to synchrotron operation, supporting closed-loop feedback systems that preserve beam stability, mitigate resonance-driven losses, and enhance extraction efficiency. In compact synchrotrons for medical proton therapy, for example in the Shanghai Advanced Proton Therapy (SAPT) facility~\cite{zhang2023sapt}, this challenge is intensified by stringent requirements: control systems require real-time tune updates while operating at signal-to-noise ratios (SNR) as low as $-20$\,dB due to the limited sensitivity and electrode length of the pickup systems, in addition to the low beam current.

The current tune measurement approach at SAPT relies on a slow-extraction kicker for beam excitation, followed by Fast Fourier Transform (FFT) analysis of beam position monitor (BPM) signals. However, the excitation interferes with the extraction process, and the residual oscillations decay rapidly after injection, making this approach impractical during slow extraction or coasting-beam operation. Under these circumstances, non-invasive diagnostics based on Schottky spectra~\cite{van2005diagnostics,van1972stochastic}, which measure incoherent betatron oscillations without perturbing the beam, provide an attractive alternative. While Schottky-based diagnostics have been successfully deployed in large hadron colliders~\cite{lasochaonline,lasocha2024jacow}, their application to compact medical synchrotrons faces severe challenges due to weak signal strength and the short sampling times available in clinical accelerators.

Conventional Schottky-based tune measurement typically employs peak-detection, curve-fitting, or mirrored-difference algorithms~\cite{betz2017bunched,lasocha2022estimation}. Peak-detection approaches locate local maxima within predefined regions, curve-fitting methods model incoherent distributions after coherent line removal, and mirrored-difference algorithms exploit sideband symmetry but require high frequency resolution and long acquisition times. Although straightforward to implement, these methods provide limited robustness at low SNR, where noise fluctuations and electronic disturbances introduce spurious peaks, and spectral broadening caused by synchrotron motion or chromaticity can obscure the true sideband structure. As a result, reliable tune extraction in real time remains difficult under the stringent performance requirements of compact proton-therapy synchrotrons.

Recent advances in machine learning for accelerator diagnostics suggest an alternative paradigm in which neural networks trained on diverse synthetic datasets can learn robust feature representations that generalize across varying beam parameters and noise characteristics~\cite{bradicic_2024_nn3sx-4wc63,MenorDeOnate:2860214}. However, existing approaches typically focus on off-line processing and do not address inference latency or provide calibrated uncertainty estimates suitable for integration with adaptive temporal filtering in feedback loops.

This paper introduces a lightweight deep learning architecture designed specifically for real-time betatron tune measurement from Schottky spectra in compact proton-therapy synchrotrons. The system addresses three key challenges: (1) achieving high accuracy under low SNR conditions, (2) maintaining sub-millisecond inference latency for integration within control systems, and (3) providing calibrated uncertainty estimates that enable uncertainty-aware temporal filtering. The main contributions are as follows:
\begin{enumerate}
    \item \textbf{Self-calibrating frequency normalization:} We propose a preprocessing pipeline that performs self-calibrating frequency normalization based on longitudinal harmonic spacing. This procedure eliminates dependence on external revolution-frequency measurements and harmonizes spectra across operating conditions by deriving the frequency ruler directly from the detected longitudinal harmonics.
    \item \textbf{Lightweight attention-based CNN architecture for tune and uncertainty estimation:} Building on this normalized representation, we design a dual-branch convolutional network with multi-scale feature extraction and attention-based pooling. The mean-prediction branch employs sharp attention for precise peak localization, while the uncertainty branch uses softer attention to assess spectral ambiguity. The resulting architecture achieves accuracy competitive with traditional methods while enabling sub-millisecond GPU inference using about $2.0\times 10^{4}$ trainable parameters.
    \item \textbf{Calibrated uncertainty quantification:} By training with a Laplace negative log-likelihood loss and employing a decoupled soft-attention mechanism for uncertainty prediction, the network learns to produce uncertainty estimates whose magnitude tracks the instantaneous prediction error. This calibration property allows the diagnostic to signal when individual frame-level predictions are likely to be unreliable.
    \item \textbf{Uncertainty-aware Kalman filtering:} We integrate the CNN outputs with a simple one-dimensional Kalman filter that interprets the predicted uncertainty as adaptive measurement noise. This temporal fusion provides substantial error reduction at low SNR levels, automatically down-weighting noisy frames without manual tuning of filter parameters.
\end{enumerate}

We validate the proposed approach on a large-scale synthetic dataset that spans realistic variations in beam parameters, including different revolution frequencies, betatron tunes, detector bandwidths, sampling rates, and SNR levels. Comprehensive evaluation demonstrates robust performance across this parameter space, with preliminary validation on operational beam data from the SAPT facility confirming successful generalization to real accelerator signals without retraining. The lightweight design enables deployment on consumer-grade hardware, addressing the practical constraints of clinical proton-therapy facilities.

The remainder of this paper is organized as follows. Section~\ref{sec:methods} presents Schottky signal formulation and problem statement, and complete system, including data preprocessing, neural network architecture, training methodology, and uncertainty-aware Kalman filtering. Section~\ref{sec:results} reports experimental results on synthetic and operational beam data. Section~\ref{sec:conclusions} concludes with a summary and future directions.

\section{Background and Methods}
\label{sec:methods}

\subsection{Schottky signals and problem formulation}
\label{subsec:schottky_signals}

Schottky signals arise from the incoherent betatron oscillations of individual particles in a circulating beam and provide a non-invasive diagnostic for betatron tune measurement. For a single particle with charge $e$ and revolution frequency $f_i$, the induced current on a pickup electrode consists of approximately delta-like pulses separated by $T_i = 1/f_i$. The Fourier spectrum of this signal exhibits harmonics at integer multiples of the revolution frequency~\cite{boussard1986schottky,chanon2016schottky,nolden2001instrumentation, lasocha2020estimation}.

\textbf{Longitudinal harmonics.}
The longitudinal component produces spectral lines at frequencies
\begin{align}
    f_n = n \cdot f_{\text{rev}},
\end{align}
where $n$ is the harmonic number and $f_{\text{rev}}$ is revolution frequency. For a beam with $N$ particles exhibiting small revolution-frequency spread, these harmonics appear as narrow peaks whose amplitude scales with $N$ and whose width reflects the momentum distribution.

\textbf{Transverse betatron sidebands.}
Betatron oscillations in the transverse plane modulate the longitudinal signal, producing sidebands around each harmonic at frequencies
\begin{align}
    f_{\pm q} = (n \pm q) \cdot f_{\text{rev}},
\end{align}
where $q \in [0,1)$ is the betatron tune. The sidebands corresponding to $n+q$ and $n-q$ are symmetric about the harmonic, and their amplitude depends on the betatron-amplitude distribution and chromaticity. For tune measurement, we exploit the fact that the sideband offset $q$ from the nearest harmonic directly encodes the fractional tune, regardless of whether the sideband appears above or below the harmonic. In the proposed diagnostic, we subsequently fold the spectrum such that the effective tune lies in $q \in [0,0.5)$.

\textbf{Bunched versus coasting beams.}
For coasting beams with continuous longitudinal particle distribution, the longitudinal harmonics and transverse sidebands exhibit smooth spectral profiles determined by the momentum and betatron-amplitude distributions. In bunched beams, the discrete longitudinal bunch structure introduces additional modulation at the synchrotron frequency $f_s$, producing bessel satellites at
\begin{align}
    f_{n,q,k} = \bigl(n \pm q \pm k f_s / f_{\text{rev}}\bigr) \cdot f_{\text{rev}},
\end{align}
for integer $k$. Despite this added complexity, the fundamental betatron-tune information remains encoded in the sideband offset from the primary harmonics, and the proposed method is designed to extract tune estimates from both beam modes without requiring explicit distinction.

\textbf{Noise model.}
The measured power spectral density (PSD) is corrupted by thermal noise from the detector electronics and impulsive interference from power supplies, RF systems, and digital electronics. We model the observed spectrum as
\begin{align}
    S_{\text{obs}}(f) = S_{\text{signal}}(f) + S_{\text{noise}}(f),
\end{align}
where $S_{\text{signal}}(f)$ represents the Schottky spectrum and $S_{\text{noise}}(f)$ includes both uniform and impulsive noise components. Our evaluation spans SNR values from $0$ to $-20$~dB, representing operational conditions in compact medical synchrotrons with limited detector sensitivity.

\textbf{Problem statement.}
Given a measured PSD $S_{\text{obs}}(f)$ acquired over a finite time window, the objective is to estimate the betatron tune $q \in [0,0.5)$ with an associated uncertainty $\sigma$ that quantifies measurement reliability. The system must operate under the following constraints: (1) sub-millisecond end-to-end latency to support real-time feedback control, (2) robust performance at SNR down to $-20$~dB, and (3) minimal dependence on auxiliary measurements such as external revolution-frequency references. These requirements motivate a data-driven approach in which a neural network learns to extract tune estimates directly from frequency-normalized spectra, with uncertainty quantification enabling adaptive temporal filtering to suppress transient measurement errors.

\subsection{Data preprocessing pipeline}
\label{subsec:preprocessing}

Raw power spectral density measurements exhibit several characteristics that complicate direct neural network processing: the betatron sidebands appear at absolute frequencies $(n \pm q) \cdot f_{\text{rev}}$ that vary with the revolution frequency, the detector response is frequency-dependent, and the dynamic range varies significantly across samples due to differences in beam intensity and noise-floor levels. We address these challenges through a two-stage preprocessing pipeline consisting of self-calibrating frequency normalization followed by robust statistical normalization.

\textbf{Self-calibrating frequency normalization.}
Rather than relying on external revolution-frequency measurements, which may be unavailable or desynchronized during energy ramping, we extract $f_{\text{rev}}$ directly from the spectrum by identifying the spacing between adjacent longitudinal harmonics. The algorithm locates prominent peaks in the low-frequency region, computes their pairwise frequency differences, and selects the most consistent spacing as the revolution-frequency estimate $\hat{f}_{\text{rev}}$. This self-calibrating approach automatically adapts to variations in $f_{\text{rev}}$ without requiring auxiliary instrumentation.

To map the absolute frequency scale onto the normalized tune axis, we employ a local interval-wise normalization that is robust to non-linearities in the harmonic spacing. Instead of applying a global scaling factor, the algorithm normalizes the frequency coordinate within each interval defined by two consecutive detected harmonic peaks $f_i$ and $f_{i+1}$. For a frequency component $f_{\text{abs}}$ falling within this interval, the local normalized coordinate is computed as
\begin{align}
    q_{\text{raw}} = \frac{f_{\text{abs}} - f_{i}}{f_{i+1} - f_{i}}.
\end{align}
This value is then folded onto the half-interval $[0, 0.5)$ via $q = \min(q_{\text{raw}}, 1 - q_{\text{raw}})$, collapsing all harmonics and their betatron sidebands onto a common tune grid. In effect, the local spacing between harmonics acts as a self-calibrating frequency ruler, eliminating the dependence on any external revolution-frequency measurement.

To handle the discrete-to-continuous mapping, we employ soft binning that distributes each frequency component across adjacent grid points with weights inversely proportional to distance, producing both a power spectrum $x[q]$ and a weight map $w[q]$ on a uniform 1024-point grid spanning $[0,0.5)$. This soft-binning approach preserves spectral resolution while providing natural anti-aliasing. The two-channel representation $\{x[q], w[q]\}$ serves as input to the neural network, with the weight map encoding the non-uniform sampling density induced by the frequency-to-tune mapping.

\textbf{Robust normalization.}
Power spectral density magnitudes can vary by several orders of magnitude across samples due to differences in beam intensity, detector gain, and noise-floor levels. Standard mean–variance normalization is unsuitable because the mean and variance are strongly influenced by spurious peaks and noise-dominated regions. We instead employ Median Absolute Deviation (MAD) normalization, which remains stable in the presence of outliers.

For each mapped spectrum $x[q]$, we compute the median baseline
\begin{align}
    b_0 &= \mathrm{median}\bigl(x[q]\bigr),
\end{align}
and the scale parameter
\begin{align}
    s_0 &= 1.4826 \cdot \mathrm{median}\!\left( \bigl|x[q] - b_0\bigr| \right),
\end{align}
where the constant $1.4826$ provides consistency with the standard deviation under Gaussian statistics. The normalized spectrum is then
\begin{align}
    z[q] &= \frac{x[q] - b_0}{s_0 + \epsilon},
\end{align}
with $\epsilon = 10^{-9}$ ensuring numerical stability. In rare cases where MAD yields a near-zero scale estimate, we fall back to an interquartile-range-based scale. This hierarchical strategy ensures robust normalization across the full range of signal conditions encountered in the dataset while preserving the linear-scale spectral structure and avoiding distortion of weak components near the noise floor.

\begin{figure}
    \centering
    \includegraphics[width=0.9\linewidth]{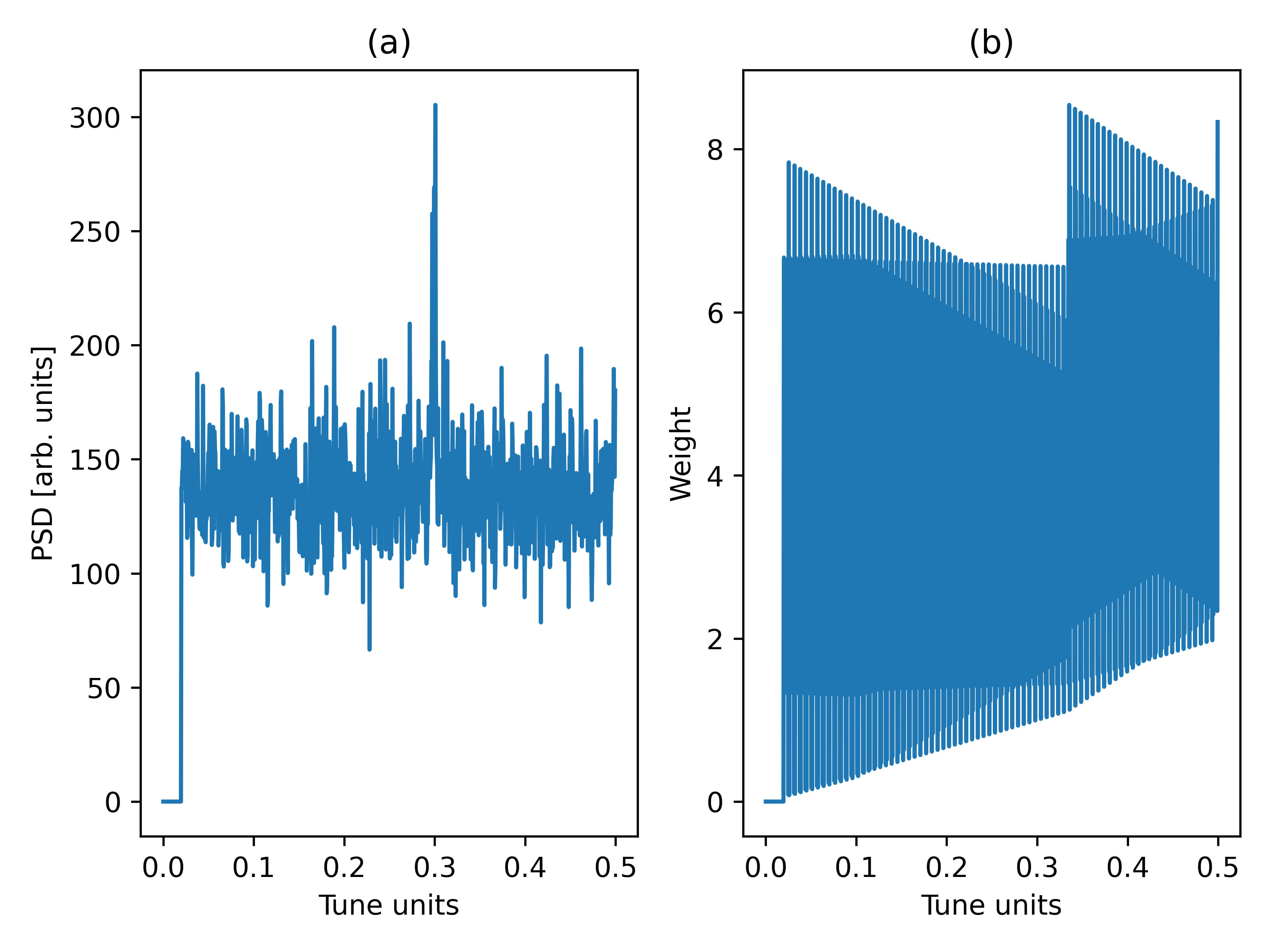}
    \caption{Mapped PSD (left) and corresponding weight map (right) after frequency normalization and soft binning. The betatron tune $q = 0.3$ appears as a localized peak in the normalized representation. SNR = $-20$ dB.}
    \label{fig:x_w}
\end{figure}

Figure~\ref{fig:x_w} shows an example of the mapped power spectrum and its corresponding weight map after preprocessing. The folding operation successfully collapses multiple harmonics onto the normalized tune axis, with the betatron sideband appearing as a localized peak at $q = 0.3$. The weight map highlights regions of dense coverage on the tune grid arising from the geometry of the frequency-to-tune mapping.

\subsection{Neural network architecture}
\label{subsec:architecture}

The neural network architecture is guided by three competing constraints inherent to real-time accelerator diagnostics: sub-millisecond inference latency to support feedback loops, robust performance under severe noise conditions with SNR values as low as $-20$ dB, and calibrated uncertainty quantification to enable adaptive temporal filtering. These requirements rule out the use of large deep learning models and motivate a lightweight architecture tailored specifically for one-dimensional tune measurement.

\textbf{Design rationale.}
Traditional peak-detection algorithms rely on argmax operations that are highly sensitive to local noise fluctuations. We replace this with soft attention pooling, where features are aggregated through learned importance weights distributed across the tune axis. This differentiable approach enables end-to-end gradient-based optimization while providing an inductive bias well-suited to one-dimensional localization. Accurate tune prediction requires precise localization focused on peak position, while reliable uncertainty estimation requires assessment of spectral ambiguity and global noise characteristics. These objectives benefit from different feature representations, motivating a dual-branch architecture that separates these pathways after shared early feature extraction. Rather than increasing network depth, which often yields limited benefit for one-dimensional spectral tasks, we employ multi-scale feature extraction through parallel fine-grained and coarse-grained convolutional pathways, concentrating parameters in channel width where they contribute more effectively to overall performance.

\textbf{Architecture overview.}
The network accepts a two-channel input tensor of shape $[2, 1024]$, where the two channels encode the normalized power spectrum and the soft-binning weight map produced by the preprocessing stage. For each input sample, the network outputs two scalar quantities: the betatron tune $q \in [0, 0.5)$ and the associated measurement uncertainty $\sigma$. The processing pipeline consists of four main stages: (1) parallel multi-scale feature extraction through fine-grained and coarse-grained convolutional stems, (2) feature fusion and shared processing via residual blocks, (3) branch splitting into specialized tune-prediction and uncertainty-estimation pathways, and (4) attention-based aggregation to produce final outputs.

\begin{figure}
    \centering
    \includegraphics[width=0.9\linewidth]{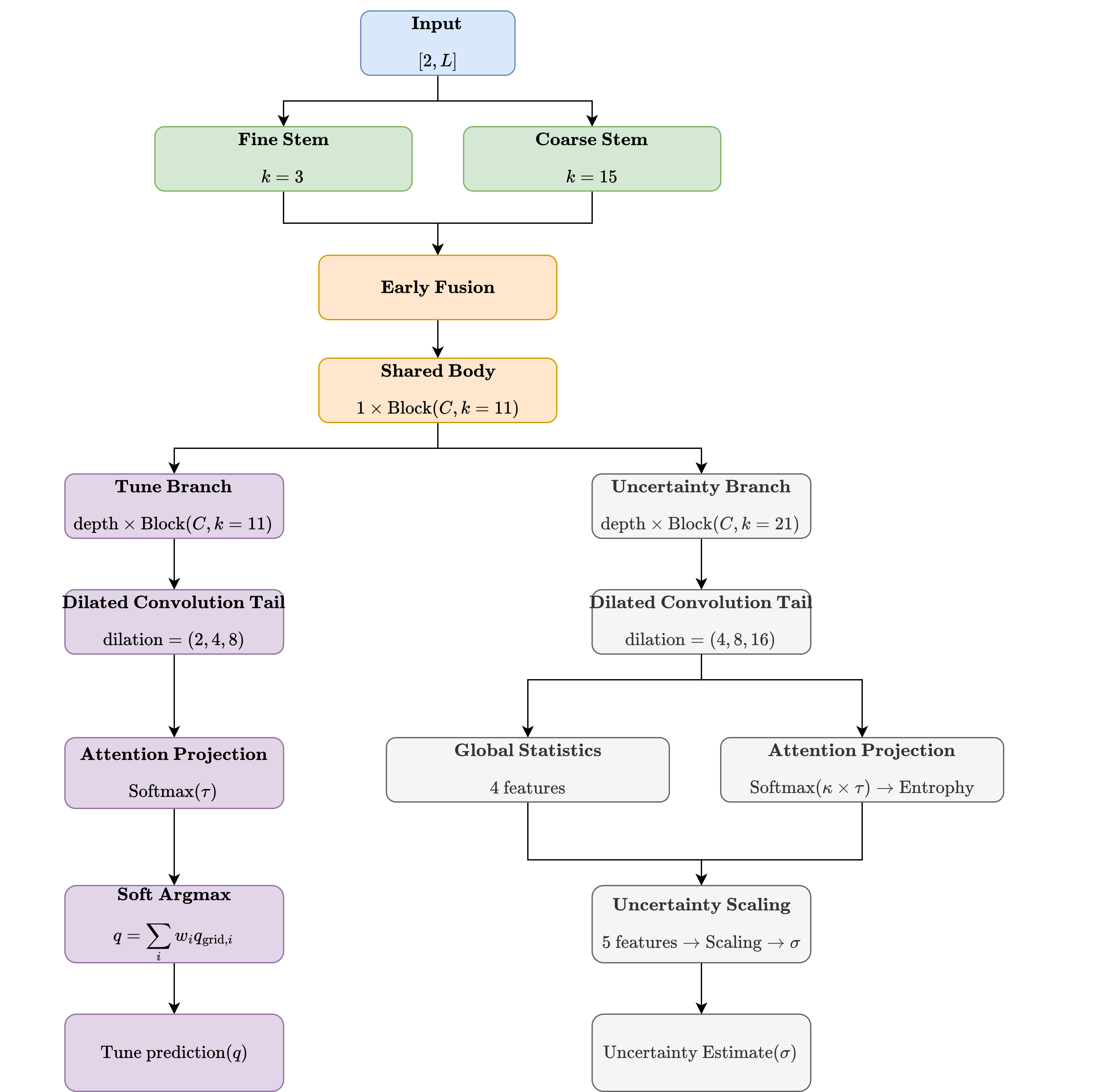}
    \caption{Overall architecture of the proposed neural network. Multi-scale feature extraction processes the two-channel input through parallel fine-grained and coarse-grained paths. After shared residual processing, the network splits into dual branches: the tune prediction branch uses sharp attention pooling for precise localization, while the uncertainty estimation branch employs softer attention to assess measurement difficulty.}
    \label{fig:overall_architecture}
\end{figure}

\begin{figure}
    \centering
    \includegraphics[width=0.9\linewidth]{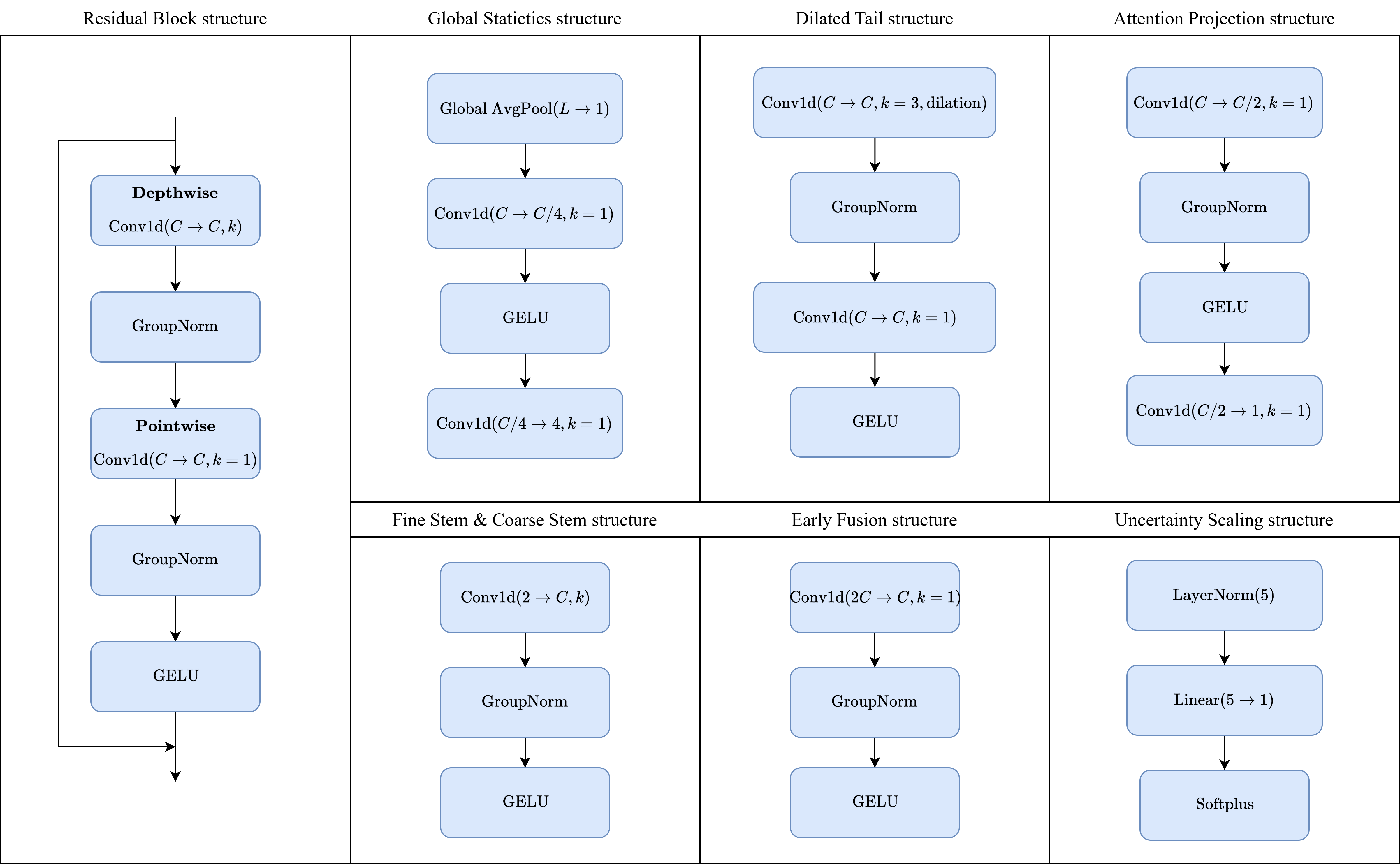}
    \caption{Detailed structure of the modules presented in Fig.~\ref{fig:overall_architecture}, where $k$ denotes the kernel size and $C$ denotes the number of channels.}
    \label{fig:detailed_architecture}
\end{figure}

Figure~\ref{fig:overall_architecture} shows the overall architecture, and the detailed structure of its constituent modules is presented in Fig.~\ref{fig:detailed_architecture}. The multi-scale stem processes the input through two parallel paths with different receptive fields: a fine-grained path with smaller kernels (sizes 3–5) captures local spectral details, while a coarse-grained path with larger kernels (sizes 7–15) captures broader contextual information. Features from both paths are concatenated and processed through a sequence of residual blocks with depthwise-separable convolutions and group normalization. After shared processing, the network splits into two specialized branches.

\textbf{Tune prediction branch.}
The tune prediction branch employs attention-based pooling over a predefined uniform grid of $L$ tune candidates spanning $[0, 0.5)$, where $L = 1024$ matches the input tune grid. Given the feature map $\mathbf{h} \in \mathbb{R}^{C \times L}$ from the shared encoder, a convolutional layer generates attention logits $\mathbf{a} \in \mathbb{R}^{M}$, which are converted to normalized weights via softmax with an adaptive temperature parameter $\tau$. The predicted tune is computed as the weighted average
\begin{align}
    \hat{q} = \sum_{i=1}^{M} \alpha_i \cdot q_i,
\end{align}
where $\alpha_i = \exp(a_i / \tau) / \sum_j \exp(a_j / \tau)$ and $q_i$ are the grid points. The temperature $\tau$ is constrained to prevent excessive sharpening that would break gradient flow, enabling the network to learn an appropriate balance between precise localization and smooth optimization.

\textbf{Uncertainty estimation branch.}
The uncertainty branch uses a similar attention mechanism but with softer weighting to capture spectral ambiguity rather than precise peak location. From the attention distribution over the tune grid, it derives a small set of difficulty features (such as the standard deviation and entropy of the weights) that quantify peak sharpness and multimodality. Global context features aggregated via adaptive average pooling are concatenated with these attention-derived features and passed through a small fully connected network to predict a positive scaling factor. The final uncertainty estimate $\sigma$ is obtained by scaling a base width computed from the attention distribution, ensuring that $\sigma$ reflects both local ambiguity around the main peak and global noise conditions.

\textbf{Model configurations.}
We evaluate three model configurations differing in base channel count $C$: a minimal 6.6k-parameter model ($C = 16$), a standard 20k-parameter model ($C = 32$), and a larger 70k-parameter model ($C = 64$). All configurations employ depthwise-separable convolutions to balance parameter efficiency with expressive capacity, with network depth fixed at 2 residual blocks. This design achieves competitive accuracy with traditional signal processing methods while maintaining the computational efficiency required for deployment on consumer-grade GPUs.

\subsection{Training and uncertainty-aware filtering}
\label{subsec:training_filtering}

\textbf{Loss function for calibrated uncertainty.}
The network is trained to jointly predict the betatron tune and its associated uncertainty using a Laplace negative log-likelihood (NLL) loss, which encourages well-calibrated uncertainty estimates while remaining robust to outliers. For a batch of $N$ samples with prediction errors $e_i = q_{\text{pred},i} - q_{\text{true},i}$ and predicted uncertainties $\sigma_i$, the loss is
\begin{align}
    \mathcal{L}_{\text{NLL}}
    &= \frac{1}{N} \sum_{i=1}^{N} \left( \frac{|e_i|}{\sigma_i} + \log \sigma_i \right).
\end{align}
When the predicted uncertainty is too small relative to the actual error, the first term grows rapidly, penalizing overconfident mispredictions; the logarithmic term prevents the network from trivially inflating $\sigma$ to reduce the error term. Compared with a Gaussian likelihood, the Laplace distribution’s heavier tails yield an MAE-like penalty that is more robust to occasional large errors at extreme SNR levels. For numerical stability and to regularize the scale of the predicted uncertainties, $\sigma$ is clipped to a fixed interval $[\epsilon, 0.2]$ with $\epsilon = 10^{-8}$.

\textbf{Optimization details.}
The network is trained using the AdamW optimizer~\cite{adamw} with an initial learning rate of $5 \times 10^{-4}$ and a weight decay of $1 \times 10^{-5}$. The learning rate is scheduled using cosine annealing over the full training duration, decreasing smoothly from the initial value to $5 \times 10^{-6}$ (1\% of the initial rate). Global gradient clipping with a maximum L2 norm of $1.0$ is applied before each optimization step to prevent occasional large gradients—amplified by temperature-scaled softmax operations in the attention modules—from destabilizing training. The model is trained for 100 epochs with a batch size of 128; each epoch comprises 1024 training steps and 512 validation steps, with each batch constructed by randomly sampling independent frames from the training set. This sampling strategy suppresses temporal correlations between consecutive batches and yields approximately i.i.d.\ gradient estimates.

Uncertainty calibration is monitored throughout training by stratifying validation samples into uncertainty quartiles and verifying that the mean and 95th-percentile errors increase monotonically from the lowest- to highest-uncertainty bin. In addition, the Spearman rank correlation between absolute errors and predicted uncertainties is evaluated on the validation set. For deployment, the final model checkpoint is selected using both prediction accuracy and calibration quality: among saved epochs with similar validation MAE, the checkpoint with higher Spearman correlation is preferred.

\textbf{Uncertainty-aware Kalman filtering.}
To exploit temporal correlations in sequential tune measurements and suppress transient outliers, the neural network predictions are combined with an uncertainty-aware Kalman filter~\cite{kendall2016modelling, kendall2017uncertainties, loquercio2020general, he2023survey}. The filter tracks the betatron tune using a scalar random-walk state model,
\begin{align}
    x_{k} &= x_{k-1} + w_{k}, && w_{k} \sim \mathcal{N}(0, Q), \\
    z_{k} &= x_{k} + v_{k},   && v_{k} \sim \mathcal{N}(0, R_{k}),
\end{align}
where $x_k \in [0, 0.5)$ denotes the true tune at time step $k$, $z_k$ is the observed measurement (the neural network prediction $q_{\text{nn},k}$), $w_k$ represents process noise with variance $Q$, and $v_k$ represents measurement noise with time-varying variance $R_k$. The state transition scalar $F = 1$ implements a first-order random walk (constant-position model), and the observation scalar $H = 1$ indicates that the tune state is measured directly.

The process noise variance is set to $Q = 10^{-6}$, which provides a practical balance between temporal smoothing and responsiveness given the slow tune drift and kilohertz-scale measurement rate typical of clinical synchrotron operation. The measurement noise variance is determined adaptively from the network’s per-frame uncertainty,
\begin{align}
    R_{k} &\approx \sigma_{\text{nn},k}^{2},
\end{align}
with a numerical floor $R_k \geq 10^{-12}$ to prevent overconfident predictions from causing instability. The posterior state variance is additionally clamped to the interval $[10^{-12}, 0.0625]$ to maintain physically meaningful uncertainty bounds consistent with the tune range $[0, 0.5)$. The Kalman recursion follows the standard predict–update form, and the filter is initialized with a broad prior ($\hat{q}_0 = 0.25$, $P_0 = 0.01^2$) to avoid biasing the initial convergence. This adaptive measurement-noise mechanism automatically down-weights frames with large predicted uncertainty—typically corresponding to low-SNR or spectrally ambiguous conditions—while trusting high-confidence predictions, yielding substantial error reduction at low SNR as quantified in Section~\ref{sec:results}.

\section{Experimental results}
\label{sec:results}

\subsection{Evaluation setup}
\label{subsec:eval_setup}

Except for the preliminary validation on operational beam data presented in Section~\ref{subsec:real_beam_validation}, all quantitative experiments are conducted offline on synthetic Schottky spectra for which ground-truth tune labels are precisely known. In all cases, whether using simulated or recorded spectra, the processing pipeline operates in a strictly causal, frame-by-frame streaming mode without access to future frames, thereby emulating an online, real-time deployment scenario.

\textbf{Static test set.}
The static performance characterization employs an independent test set of 50,000 spectra stratified uniformly across SNR levels from $0$ to $-20$~dB, detector bandwidths from 8 to 26~MHz, revolution frequencies from 4 to 7.5~MHz (covering the energy range of SAPT), and both bunched and coasting beam modes. The betatron tune values are sampled uniformly from the operational range $q \in [0.05, 0.45]$, excluding regions near integer and half-integer resonances. This broad parameter coverage ensures that the reported performance metrics reflect robustness across the diverse operating conditions encountered during clinical beam delivery, including injection plateaus, energy ramping phases, and extraction processes.

\textbf{Temporal test sequences.}
The temporal validation experiments utilize three independent sequences, each comprising 10,000 consecutive frames generated with smoothly varying betatron tune and revolution frequency. The three sequences are configured at SNR levels of $-20$, $-15$, and $-10$~dB, spanning conditions from severely degraded measurements dominated by noise to typical clinical operating conditions. All temporal sequences employ a fixed detector bandwidth of 8~MHz, representing the most challenging preprocessing scenario with minimal frequency averaging per tune bin and ensuring that at least two longitudinal harmonics are present within the analyzed band.

The chosen SNR range ($0$ to $-20$~dB) reflects measurements from clinical compact synchrotrons operating with pickup systems whose electrode length is constrained by ring-circumference limitations, where the typical SNR during stable beam delivery lies between $-10$ and $-15$~dB and can degrade to around $-20$~dB in worst-case conditions. The bandwidth range (8--26~MHz) corresponds to the frequency region in which the pickup structure exhibits relatively high quality factor. The revolution-frequency range (4--7.5~MHz) spans the operational envelope of SAPT, covering extraction energies from approximately 70 to 230~MeV.

\textbf{Performance metrics.}
Evaluation employs complementary metrics that assess both prediction accuracy and uncertainty calibration quality. The primary accuracy metric is the mean absolute error (MAE):
\begin{align}
    \text{MAE} = \frac{1}{N} \sum_{i=1}^{N} \bigl|q_{\text{pred},i} - q_{\text{true},i}\bigr|.
\end{align}
We additionally report the 95th-percentile error (P95), which captures tail performance critical for safety-critical beam control applications, and the fraction of predictions achieving high accuracy:
\begin{align}
    P(<10^{-3}) = \frac{1}{N} \sum_{i=1}^{N} \mathbb{1}\bigl(|e_i| < 10^{-3}\bigr),
\end{align}
where $e_i = q_{\text{pred},i} - q_{\text{true},i}$ is the prediction error for sample $i$. The threshold $|e| < 10^{-3}$ is particularly relevant for precision beam delivery in intensity-modulated treatments.

For uncertainty calibration, we employ the Spearman rank correlation coefficient $\rho$ between the absolute prediction errors $\lvert e_i\rvert$ and the predicted uncertainties $\sigma_{\text{pred},i}$ as a global measure of monotonic association. We interpret $\rho$ using established thresholds: $\rho < 0.3$ indicates poor calibration requiring intervention, $0.3 \leq \rho < 0.5$ suggests adequate calibration, and $\rho \geq 0.5$ indicates strong calibration sufficient for safety-critical deployment. We also perform quartile-based calibration analysis by stratifying predictions into four equal-sized groups based on predicted uncertainty magnitude and computing the mean absolute error within each quartile. In a well-calibrated model, the quartile-wise errors increase monotonically with predicted uncertainty.

\subsection{Static performance and uncertainty calibration}
\label{subsec:static_performance}

This subsection quantifies the per-frame prediction accuracy and uncertainty calibration of the neural network across the full range of operating conditions relevant to SAPT. The evaluation uses the 50,000-sample static test set distributed uniformly across SNR levels, detector bandwidths, and beam configurations.

\textbf{Performance across SNR levels.}
Table~\ref{tab:static_snr} reports the per-frame accuracy metrics as a function of SNR. The neural network maintains a mean absolute error (MAE) below $3 \times 10^{-4}$ across SNR levels down to approximately $-15~\mathrm{dB}$, demonstrating robust performance under typical clinical operating conditions. In the lowest-SNR regime between $-15$ and $-20~\mathrm{dB}$, the performance degrades to an MAE of approximately $8 \times 10^{-4}$, with the proportion of predictions satisfying $\lvert e \rvert < 10^{-3}$ remaining above $87\%$. The standard deviation of the per-frame error increases by nearly an order of magnitude in this regime, highlighting the strongly bimodal error distribution characteristic of low-SNR conditions in which most frames yield accurate predictions but occasional frames produce large errors under unfavorable noise realizations.

\begin{table}
\centering
\caption{Single-frame CNN performance across SNR levels on the static test set (50,000 samples). All measurements are computed on mixed-bandwidth (8--26~MHz) and mixed beam-mode (bunched/coasting) distributions.}
\label{tab:static_snr}
\begin{tabular}{ccccc}
\hline
SNR Range (dB) & MAE~($10^{-4}$) & $\sigma$(error)~($10^{-4}$) & P95~($10^{-4}$) & P($<10^{-3}$) \\
\hline
$-5$ to $0$     & $1.66$ & $1.63$ & $4.62$ & 99.68\% \\
$-10$ to $-5$   & $1.95$ & $1.87$ & $5.57$ & 99.53\% \\
$-15$ to $-10$  & $2.80$ & $8.72$ & $8.10$ & 97.09\% \\
$-20$ to $-15$  & $8.06$ & $83.49$ & $16.92$ & 87.00\% \\
\hline
\end{tabular}
\end{table}

The network maintains broadly consistent accuracy across the full bandwidth range at SNR $\geq -10~\mathrm{dB}$, with MAE differing by at most a factor of about $1.5$ between narrow-bandwidth (8--10~MHz) and wide-bandwidth (24--26~MHz) configurations, demonstrating that the self-calibrating frequency-normalization procedure substantially suppresses systematic errors. In the lowest-SNR regime, a strong bandwidth dependence emerges, with narrow-bandwidth configurations exhibiting MAE larger by about a factor of five compared with wide-bandwidth, reflecting reduced noise averaging. At moderate-to-high SNR ($-10$ to $0~\mathrm{dB}$), the network attains similar accuracy on both bunched and coasting beam modes, with the coasting-beam spectra yielding noticeably smaller errors than bunched beams only in the lowest-SNR regime.

\begin{figure}
    \centering
    \includegraphics[width=0.9\linewidth]{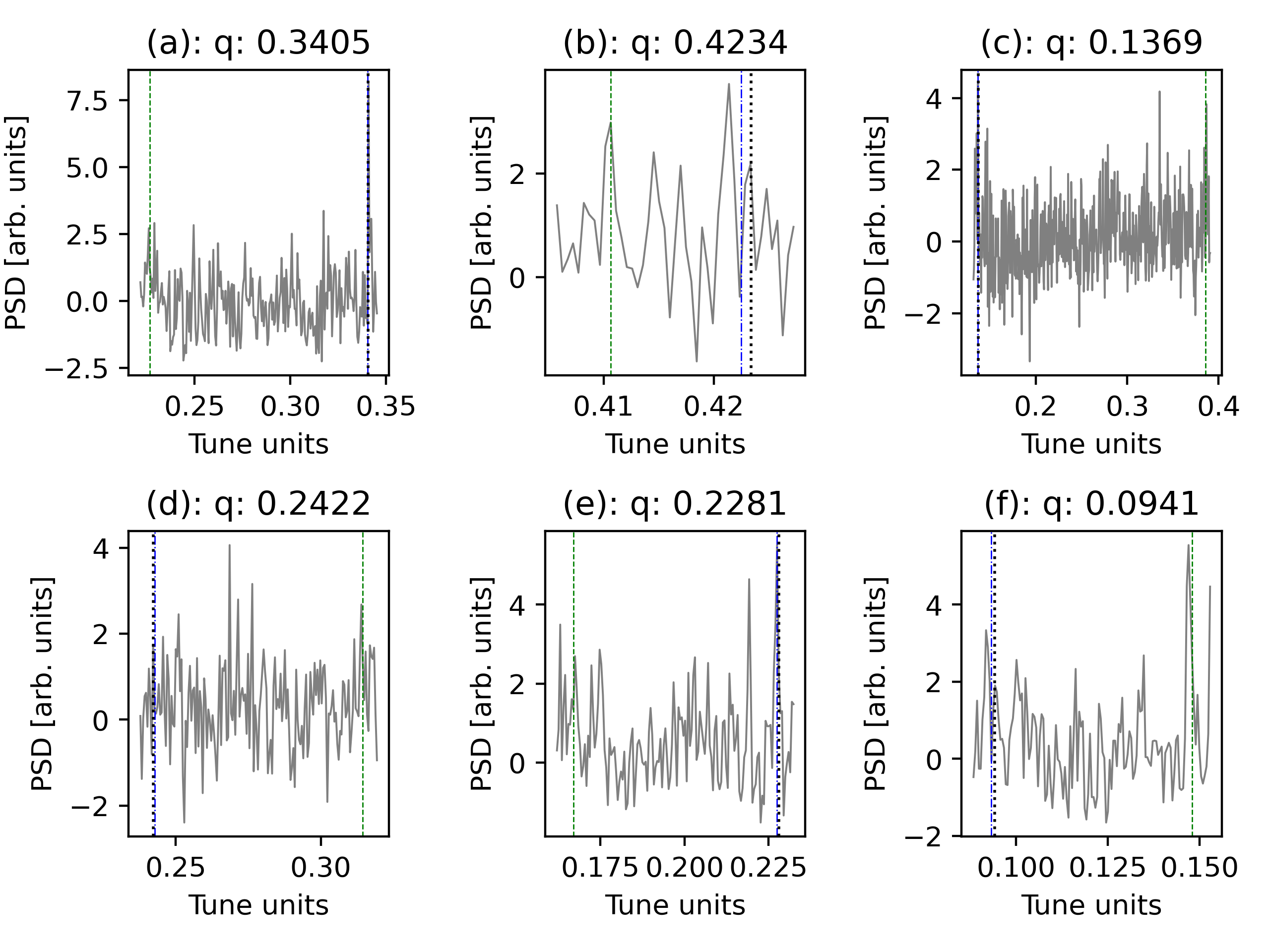}
    \caption{Failure cases for the peak-detection baseline on simulated Schottky spectra from the static test set. Each panel shows a PSD snapshot in tune units. Vertical lines indicate the true tune (black dotted line), the peak-detection estimate (green dashed line), and the CNN prediction (blue dash-dotted line). The six examples are selected such that the peak-detection error is larger than $10^{-2}$, while the CNN prediction error is below $10^{-3}$ with predicted uncertainty $\sigma < 3\times 10^{-3}$, highlighting typical low-SNR failure modes of the classical algorithm.}
    \label{fig:failure_gallery}
\end{figure}

\textbf{Qualitative failure-case analysis.}
To complement the aggregate statistics in Table~\ref{tab:static_snr}, Fig.~\ref{fig:failure_gallery} presents a gallery of representative failure cases for the peak-detection baseline on the static test set. Each panel shows a PSD snapshot in tune units around the horizontal betatron sideband, with vertical lines indicating the true tune (black), the peak-detection estimate (green dashed), and the CNN prediction (blue dotted). The six spectra are selected from the lowest-SNR range such that the peak-detection error exceeds $10^{-2}$ in tune units, whereas the CNN error remains below $10^{-3}$ with predicted uncertainty $\sigma < 3\times 10^{-3}$. These examples illustrate typical situations in which the classical algorithm locks onto spurious noise spikes or parasitic spectral structures, while the proposed CNN still identifies the correct working point with moderate confidence.

\textbf{Uncertainty calibration.}
Table~\ref{tab:calibration_spearman} reports the Spearman rank-correlation coefficient, $\rho$, between the absolute prediction errors $|e_i|$ and the predicted uncertainties $\sigma_{\text{pred},i}$. The global correlation over all 50,000 test samples is $\rho = 0.501$, indicating strong overall association. To isolate the network's learned calibration capability from average SNR effects, we compute Spearman correlations within narrow SNR ranges. The per-SNR correlations reveal that calibration quality is strongest precisely where it is most needed: at the most challenging SNR range ($-20$ to $-15~\mathrm{dB}$), the within-range correlation reaches $\rho = 0.534$, indicating successful discrimination between individual frames. As SNR increases, the per-SNR correlation decreases to $\rho = 0.365$ at excellent SNR ($-5$ to $0~\mathrm{dB}$), which is expected and desirable as the error distribution becomes increasingly concentrated near zero. All per-SNR correlations remain at or above the adequate calibration threshold ($\rho \geq 0.3$), confirming that the uncertainty estimates provide meaningful information even within homogeneous SNR regimes.

\begin{table}
\centering
\caption{Spearman rank-correlation coefficient $\rho$ between absolute prediction errors and predicted uncertainties. The global correlation is computed over all 50,000 test samples; per-SNR correlations are computed within each SNR range to isolate calibration quality from SNR-driven error variation.}
\label{tab:calibration_spearman}
\begin{tabular}{cc}
\hline
SNR range (dB) & Spearman $\rho$ \\
\hline
\textit{All SNRs} ($-20$ to $0$) & 0.501 \\
\hline
$-5$ to $0$     & 0.365 \\
$-10$ to $-5$   & 0.400 \\
$-15$ to $-10$  & 0.450 \\
$-20$ to $-15$  & 0.534 \\
\hline
\end{tabular}
\end{table}

To assess whether the network's uncertainty estimates provide practically useful stratification of measurement reliability, we partition the test set into four equal-size groups (quartiles) based on the predicted uncertainty magnitude and compute accuracy metrics within each quartile. Figure~\ref{fig:calibration_quartiles} demonstrates strict adherence to the calibration ordering criterion across all tested SNR ranges, with both MAE and P95 increasing monotonically from Q1 to Q4 in every case. At the most challenging SNR range ($-20$ to $-15~\mathrm{dB}$), the Q4 MAE exceeds the Q1 MAE by nearly an order of magnitude. When aggregating all SNR levels, the global quartile split remains strong, with Q4 MAE being about eight times larger than Q1, confirming that the uncertainty-based ranking remains informative even when pooling heterogeneous operating conditions. The P95 error exhibits even stronger quartile separation than MAE, with Q4 tail errors on the order of a few $10^{-3}$ at low SNR, while Q1 tail errors remain at the level of a few $10^{-4}$.

\begin{figure}
\centering
\includegraphics[width=0.9\textwidth]{calibration_quartiles.png}
\caption{Mean absolute error within each uncertainty quartile (Q1--Q4) as a function of SNR. The strict monotonic ordering (Q1 $<$ Q2 $<$ Q3 $<$ Q4) at all SNR ranges confirms robust calibration across operating conditions.}
\label{fig:calibration_quartiles}
\end{figure}

The demonstrated calibration quality provides a solid foundation for effective uncertainty-aware Kalman filtering. By using the predicted uncertainties to set an adaptive measurement noise variance $R_k = \max\!\bigl(\sigma_{\text{pred},k}^2, R_{\min}\bigr)$, the Kalman filter automatically assigns low weight to Q4 frames with characteristically large errors while placing high weight on Q1 frames with high reliability, enabling the filter to suppress outliers without manual threshold tuning.

\textbf{Comparison with traditional peak detection.}
Table~\ref{tab:comparison_peak_detection} compares the accuracy of the CNN and a traditional peak-detection baseline applied to the same test spectra. The baseline implements a classical tune-measurement algorithm: the mapped tune-axis spectrum is smoothed to suppress high-frequency noise, and the dominant betatron peak is identified by prominence-based peak detection. The neural network achieves comparable or superior performance across the full SNR range, with its advantage over peak detection becoming progressively more pronounced as signal quality degrades. At high SNR ($-5$ to $0~\mathrm{dB}$), both methods perform excellently with MAE below $5 \times 10^{-4}$. For SNR in the range $-15$ to $-10~\mathrm{dB}$, the CNN achieves an MAE of $2.81 \times 10^{-4}$ compared with $4.00 \times 10^{-4}$ for peak detection, corresponding to an error reduction of approximately 30\%. In the lowest-SNR range between $-20$ and $-15~\mathrm{dB}$, the CNN achieves an MAE of $7.99 \times 10^{-4}$ versus $1.172 \times 10^{-3}$ for peak detection, corresponding to an error reduction of about 32\%.

\begin{table}
\centering
\caption{Performance comparison between CNN and traditional peak detection across SNR ranges on the static test set. Both methods process identical frequency-normalized spectra.}
\label{tab:comparison_peak_detection}
\begin{tabular}{ccccccc}
\hline
\multirow{2}{*}{SNR Range (dB)} & \multicolumn{3}{c}{CNN ($10^{-4}$)} & \multicolumn{3}{c}{Peak Detection ($10^{-4}$)} \\
\cline{2-7}
 & MAE & P95 & P($<10^{-3}$) & MAE & P95 & P($<10^{-3}$) \\
\hline
$-5$ to $0$     & 1.66 & 4.63 & 99.69\% & 3.25 & 9.88 & 95.13\% \\
$-10$ to $-5$   & 1.95 & 5.59 & 99.53\% & 3.41 & 10.31 & 94.64\% \\
$-15$ to $-10$  & 2.81 & 8.11 & 97.13\% & 4.00 & 12.34 & 92.16\% \\
$-20$ to $-15$  & 7.99 & 16.91 & 87.06\% & 11.72 & 21.61 & 83.77\% \\
\hline
\end{tabular}
\end{table}

\subsection{Temporal tracking performance}
\label{subsec:temporal_performance}

The static performance characterization evaluated single-frame prediction accuracy on independently sampled spectra. Real-time beam diagnostics, however, operate on continuous temporal sequences, enabling temporal filtering strategies that exploit correlations between consecutive frames to suppress noise and improve tracking robustness. This subsection evaluates the performance of uncertainty-aware Kalman filtering on synthetic temporal sequences.

\begin{figure}
\centering
\includegraphics[width=0.9\linewidth]{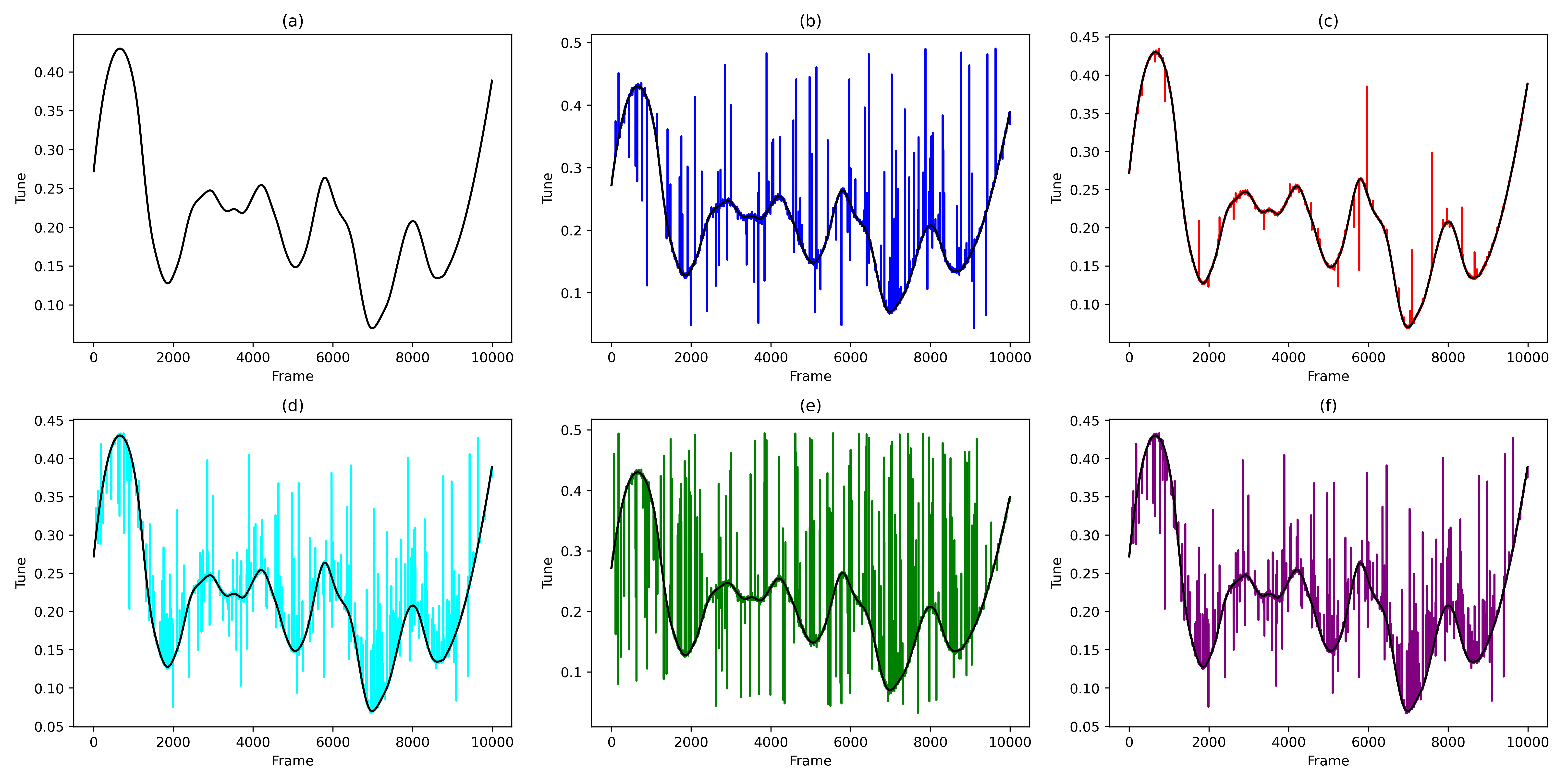}
\caption{Comparison of tracking trajectories at SNR = $-20$~dB. (a) Ground-truth tune trajectory. (b) Raw neural-network predictions. (c) Proposed uncertainty-aware method. (d) Neural-network with fixed-R Kalman filter. (e) Peak-detection baseline. (f) Peak-detection with fixed-R Kalman filter. The proposed method exhibits more stable tracking with fewer outliers. It should be noted that the tune trajectory is used solely for evaluating tracking performance and does not reflect the true tune variation during operation.}
\label{fig:approach_comparison}
\end{figure}

We construct three independent temporal sequences, each comprising 10,000 consecutive frames, configured at SNR levels of $-20$ and $-15$\,dB. All sequences employ a fixed detector bandwidth of 8\,MHz and a smoothly varying revolution frequency ranging from 4.0 to 7.5\,MHz, representing a challenging preprocessing scenario. The comparison of tracking trajectories at SNR = $-20$~dB is presented in Fig.~\ref{fig:approach_comparison}. Within each sequence, the true betatron tune follows a random smooth trajectory designed solely for evaluation purposes, with consecutive frames exhibiting small tune increments on the order of $1.5 \times 10^{-4}$.

\textbf{Comparative evaluation of filtering strategies.}
Table~\ref{tab:temporal_comparison} presents a comprehensive comparison of temporal tracking performance across five filtering configurations: (1) traditional peak detection without filtering, (2) peak detection with fixed-variance Kalman smoothing, (3) neural-network predictions without filtering, (4) neural-network predictions with fixed-variance Kalman filtering, and (5) neural-network predictions with uncertainty-aware adaptive Kalman filtering (the proposed system). For fixed-variance Kalman filtering, the measurement-noise parameter $R$ is calibrated from validation-set uncertainty statistics at each SNR level: $R = (4 \times 10^{-4})^2$ at $-15$\,dB and $R = (1.5 \times 10^{-3})^2$ at $-20$\,dB. The proposed uncertainty-aware system uses the same process noise $Q = 10^{-6}$ but sets the measurement noise adaptively as $R_k = \sigma_{\text{nn},k}^2$ based on the neural network's frame-specific uncertainty predictions.

\begin{table}
\centering
\caption{Comprehensive temporal tracking performance comparison across filtering strategies at SNR = $-15$\,dB and $-20$\,dB. All methods process identical 10,000-frame synthetic sequences with detector bandwidth 8\,MHz and revolution frequency $f_{\text{rev}}$ varying smoothly within 4.0–7.5\,MHz. Fixed-R Kalman uses validation-calibrated measurement-noise variance; Adaptive-R Kalman (proposed) uses frame-specific $R_k = \sigma_{\text{nn},k}^2$.}
\label{tab:temporal_comparison}
\begin{tabular}{lcccccc}
\hline
\multirow{2}{*}{Method} & \multicolumn{3}{c}{SNR = $-15$ dB ($10^{-4}$)} & \multicolumn{3}{c}{SNR = $-20$ dB ($10^{-4}$)} \\
\cline{2-7}
 & MAE & P95 & P($<10^{-3}$) & MAE & P95 & P($<10^{-3}$) \\
\hline
Peak Detection                 & 4.85 & 15.51 & 88.79\% & 45.07 & 37.58 & 73.72\% \\
Peak Detection + Fixed-R KF    & 4.59 & 14.38 & 89.80\% & 35.23 & 38.17 & 74.70\% \\
\hline
CNN                            & 3.69 & 11.13 & 93.63\% & 22.24 & 31.03 & 74.88\% \\
CNN + Fixed-R KF               & 3.54 & 10.88 & 93.92\% & 20.66 & 32.62 & 76.26\% \\
CNN + Adaptive-R KF (Proposed) & 2.90 &  8.18 & 97.54\% &  6.40 & 15.13 & 87.53\% \\
\hline
\end{tabular}
\end{table}

The results reveal three key findings. First, neural-network predictions consistently outperform traditional peak detection: the CNN achieves approximately 25\% lower MAE than peak detection at $-15$\,dB and roughly 50\% lower MAE at $-20$\,dB without temporal filtering. Second, fixed-variance Kalman filtering provides modest improvement, reducing MAE by only about 4\% at $-15$\,dB and 7\% at $-20$\,dB for the CNN. The limited improvement occurs because the fixed measurement-noise parameter represents an average over validation data and cannot capture the heterogeneous frame-to-frame measurement quality characteristic of low-SNR operation.

Third, uncertainty-aware adaptive filtering substantially enhances performance, particularly at the most challenging SNR conditions. At $-15$\,dB, adaptive filtering reduces MAE by about 18\% relative to fixed-variance filtering, achieving $2.90 \times 10^{-4}$. At $-20$\,dB, the advantage becomes pronounced, with adaptive filtering achieving an MAE of $6.40 \times 10^{-4}$, which represents approximately a 69\% reduction compared to fixed-variance filtering and about a 71\% improvement over the unfiltered CNN predictions. The P95 error improvement is even more pronounced, with the adaptive system achieving $1.51 \times 10^{-3}$ at $-20$\,dB compared to $3.26 \times 10^{-3}$ for fixed-variance filtering.

The superior performance of uncertainty-aware filtering stems from its ability to modulate trust in individual measurements based on spectral quality. At $-20$\,dB SNR, the neural network's predicted uncertainties span more than three orders of magnitude across frames, reflecting genuine variation in measurement difficulty as noise realizations alternately enhance or corrupt betatron peak visibility. The adaptive measurement noise $R_k = \sigma_{\text{nn},k}^2$ translates this uncertainty information into Kalman-gain modulation: clean frames with low predicted uncertainty yield high Kalman gain (the filter trusts the measurement), whereas corrupted frames with high predicted uncertainty yield low gain (the filter relies on the process-model prediction). This adaptive weighting implements a data-driven fusion strategy that neither discards potentially informative measurements nor blindly trusts unreliable observations.

\textbf{Performance summary across SNR levels.}
At excellent SNR (around $-10$\,dB), the system achieves MAE below $2 \times 10^{-4}$ with more than 99\% of frames meeting the high-accuracy criterion ($< 10^{-3}$). At lower SNR ($-15$\,dB), adaptive Kalman filtering achieves MAE $2.90 \times 10^{-4}$, providing reliable tracking suitable for routine beam delivery and closed-loop feedback. At challenging conditions ($-20$\,dB SNR) representative of severely bandwidth-constrained systems, the system achieves MAE $6.40 \times 10^{-4}$ with 87.5\% high-accuracy predictions. All configurations tested achieve typical tune-measurement errors well within clinical accuracy requirements ($\pm 10^{-3}$) at SNR $\geq -15$\,dB.

\subsection{Ablation studies and system characterization}
\label{subsec:ablation}

\textbf{Effect of loss function on uncertainty calibration.}
The proposed system employs a Laplace negative log-likelihood (NLL) loss for training, which jointly supervises tune prediction accuracy and uncertainty calibration. To evaluate whether NLL-based training maintains competitive prediction accuracy while enabling calibrated uncertainty estimates, we compare two identical network instances (approximately 20k parameters, dual-branch design) trained on the same dataset using either MAE loss or Laplace NLL loss.

Table~\ref{tab:ablation_loss} presents the static test-set performance for both training configurations. The MAE-trained and NLL-trained networks achieve very similar prediction accuracy, with MAE differing by on the order of 10\% and P95 error being virtually identical. This similarity confirms that the additional uncertainty supervision in the Laplace NLL loss does not materially compromise prediction quality. However, the training objectives produce dramatically different uncertainty calibration quality. The NLL-trained network achieves a Spearman rank correlation of approximately $\rho = 0.50$ and maintains strict monotonic quartile ordering. The MAE-trained network, despite possessing an identical architectural uncertainty branch, produces poorly calibrated uncertainties with $\rho \approx -0.09$ and violates the expected monotonic quartile ordering at multiple SNR levels. This degradation occurs because the MAE loss provides no gradient signal to the uncertainty head, resulting in uncertainty predictions that convey no meaningful information about measurement reliability.

\begin{table}
\centering
\caption{Static test-set performance comparison between MAE-trained and NLL-trained networks. Both employ an identical architecture (approximately 20k parameters, dual-branch design).}
\label{tab:ablation_loss}
\begin{tabular}{lcccc}
\hline
\multirow{2}{*}{Training Loss} & \multicolumn{2}{c}{Accuracy Metrics ($10^{-4}$)} & \multicolumn{2}{c}{Calibration Metrics} \\
\cline{2-5}
 & MAE & P95 & Spearman $\rho$ & Q1 < Q2 < Q3 < Q4 \\
\hline
MAE         & 3.28 & 9.08 & -0.092 & Violated  \\
Laplace NLL & 3.60 & 9.09 &  0.502 & Satisfied \\
\hline
\end{tabular}
\end{table}

The poor calibration of MAE-trained uncertainties renders them unsuitable for adaptive measurement-noise estimation in Kalman filtering. The NLL-trained network's calibrated uncertainties enable uncertainty-aware adaptive filtering that, at the most challenging SNR of $-20$\,dB, achieves an MAE reduction of approximately 70\% relative to both the unfiltered CNN and the fixed-variance Kalman baseline. This dramatic performance difference, which arises directly from the choice of training loss while maintaining an identical network architecture and similar prediction accuracy, indicates that Laplace NLL training is practically essential for the proposed uncertainty-aware diagnostic system.

\textbf{Model size and parameter efficiency.}
We compare three network configurations with varying capacity: a minimal model (6,598 parameters), the proposed baseline (20,470 parameters), and an expanded model (70,102 parameters), all employing an identical dual-branch architecture but differing in convolutional channel counts. Table~\ref{tab:ablation_size} presents the static test-set performance and single-frame inference latency. The results reveal diminishing returns beyond the proposed baseline configuration: the small 6.6k-parameter model achieves slightly lower MAE than the baseline (about 9\% reduction), whereas the large 70k-parameter model achieves only about 7\% lower MAE despite a 3.4-fold increase in parameter count. All three models maintain strong uncertainty calibration (Spearman $\rho > 0.45$), indicating that calibration quality is primarily determined by the training loss and architectural design rather than by raw model capacity.

\begin{table}
\centering
\caption{Performance comparison across model sizes on the static test set. All models employ identical dual-branch architecture with attention-based pooling.}
\label{tab:ablation_size}
\begin{tabular}{lccccc}
\hline
Model Size & Parameters & MAE ($10^{-4}$) & P95 ($10^{-4}$) & Spearman $\rho$ & Inference (ms) \\
\hline
Small              & 6{,}598   & 3.29 & 9.25 & 0.494 & 0.490 \\
Baseline (Proposed) & 20{,}470 & 3.60 & 9.09 & 0.501 & 0.500 \\
Large              & 70{,}102  & 3.36 & 8.99 & 0.511 & 0.515 \\
\hline
\end{tabular}
\end{table}

The inference latency increases only slightly with model size and remains well below the 1\,ms threshold required for real-time control. The baseline configuration represents a favorable trade-off: it achieves strong uncertainty calibration (Spearman $\rho > 0.5$), maintains tail performance within 1\% of the large model, and preserves minimal inference latency, while requiring 70\% fewer parameters than the large model.

\textbf{Real-time latency characterization.}
The proposed system achieves an end-to-end processing latency of approximately 0.90\,ms per frame on representative hardware (NVIDIA RTX 4070 Ti GPU, Intel Core i5-13490F CPU), which comfortably meets the sub-millisecond requirement for real-time beam diagnostics. The processing pipeline comprises CPU-based preprocessing (0.30\,ms), GPU-accelerated neural-network inference (0.50\,ms), and CPU-based Kalman filtering (0.05\,ms). This sub-millisecond latency without median filtering enables deployment in latency-critical applications, with approximately 0.10–0.15\,ms of margin for additional system overhead. When optional median filtering is enabled ($W = 3$–5), the method introduces a causal delay of approximately 1–2 frames, which is acceptable within typical 3–5\,ms control-loop periods.

\subsection{Preliminary validation on operational beam data}
\label{subsec:real_beam_validation}

While the preceding subsections provide comprehensive performance characterization on synthetic data, validation on authentic accelerator signals is essential to confirm that simulation-based training generalizes to real-world conditions. This subsection presents preliminary results from a limited dataset acquired during beam extraction at the SAPT facility, demonstrating that the proposed system, trained entirely on simulated spectra without fine-tuning or domain adaptation, successfully tracks the betatron tune on operational beam data.

The validation dataset comprises 625 consecutive averaged PSD frames acquired at a sampling rate of 29.75~MHz during the extraction sequence. During this period, the proton beam had a kinetic energy of 141.736~MeV and a revolution frequency of 6.04~MHz. During third-order resonance slow extraction, the horizontal betatron tune is deliberately maintained near $q \approx 0.67$. The spectra exhibit characteristics typical of clinical compact-synchrotron diagnostics: moderate SNR (estimated at $-10$ to $-15$~dB), a limited measurement bandwidth of 0–11~MHz, and persistent spurious peaks arising from RF pickup, power-supply ripple, and digital electronics.

\begin{figure}
\centering
\includegraphics[width=0.9\linewidth]{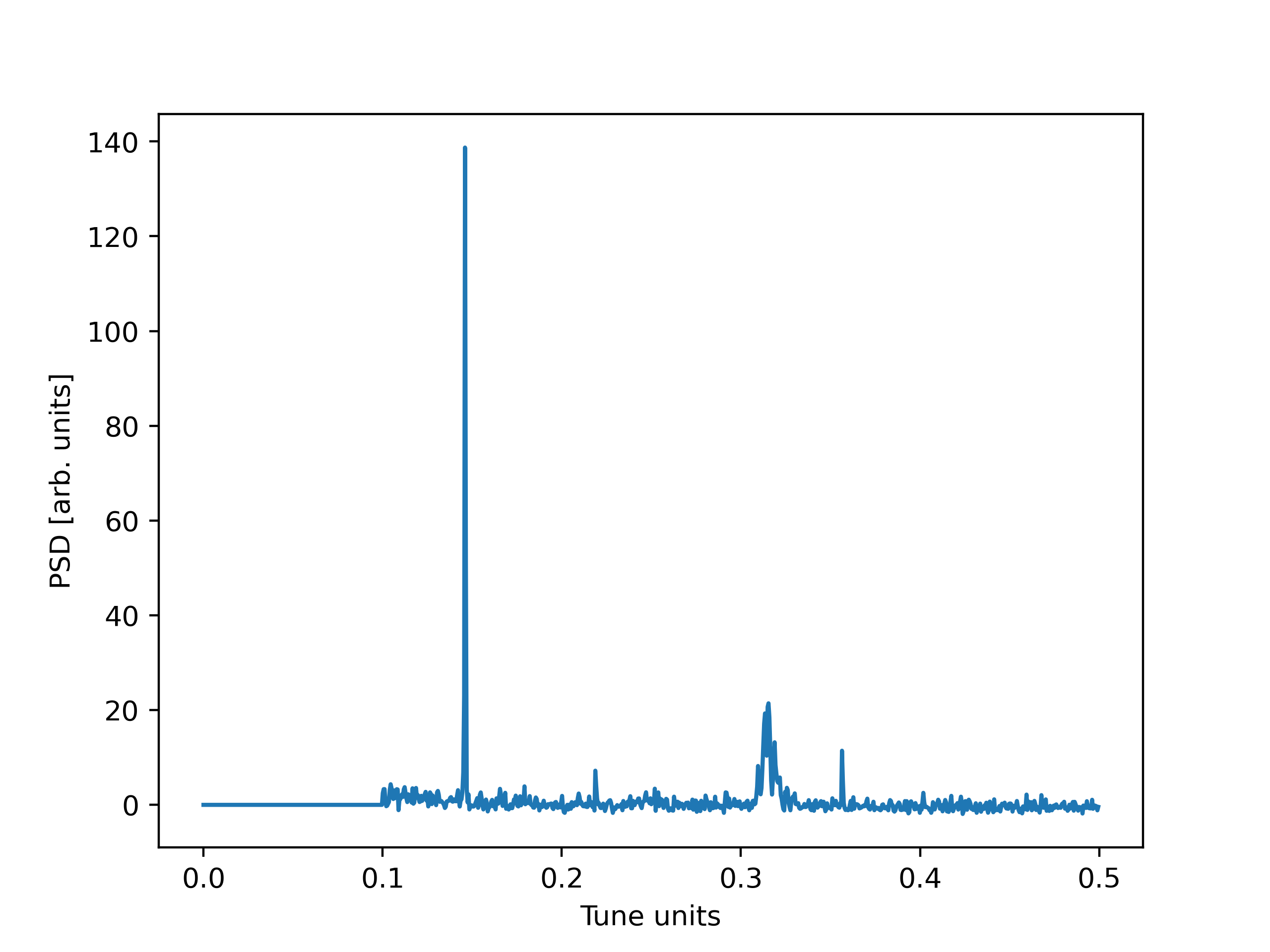}
\caption{Single-frame horizontal Schottky spectrum after mapping to the normalized tune axis $q$. A narrow, high-amplitude peak near $q = 0.14$ is identified as spurious interference, whereas the broader transverse Schottky sideband around $q_{\text{ref}} = 0.3157$ corresponds to the true horizontal betatron signal.}
\label{fig:real_spectrum}
\end{figure}

To establish ground truth for validation, all 625 spectra were averaged to suppress uncorrelated noise fluctuations, yielding an ensemble-averaged spectrum suitable for high-confidence peak identification. Conventional peak-detection analysis applied to this averaged spectrum identifies the betatron tune as $q_{\text{ref}} = 0.3157$, which serves as the reference value for evaluating frame-by-frame tracking accuracy. Figure~\ref{fig:real_spectrum} shows a representative single-frame spectrum exhibiting multiple narrow interference lines at distinct tune values, with amplitudes that can be either larger or smaller than those of the transverse betatron sidebands. Conventional peak-detection algorithms may erroneously lock onto an interference line when its instantaneous amplitude fluctuates upward.

\begin{figure}
\centering
\includegraphics[width=0.9\textwidth]{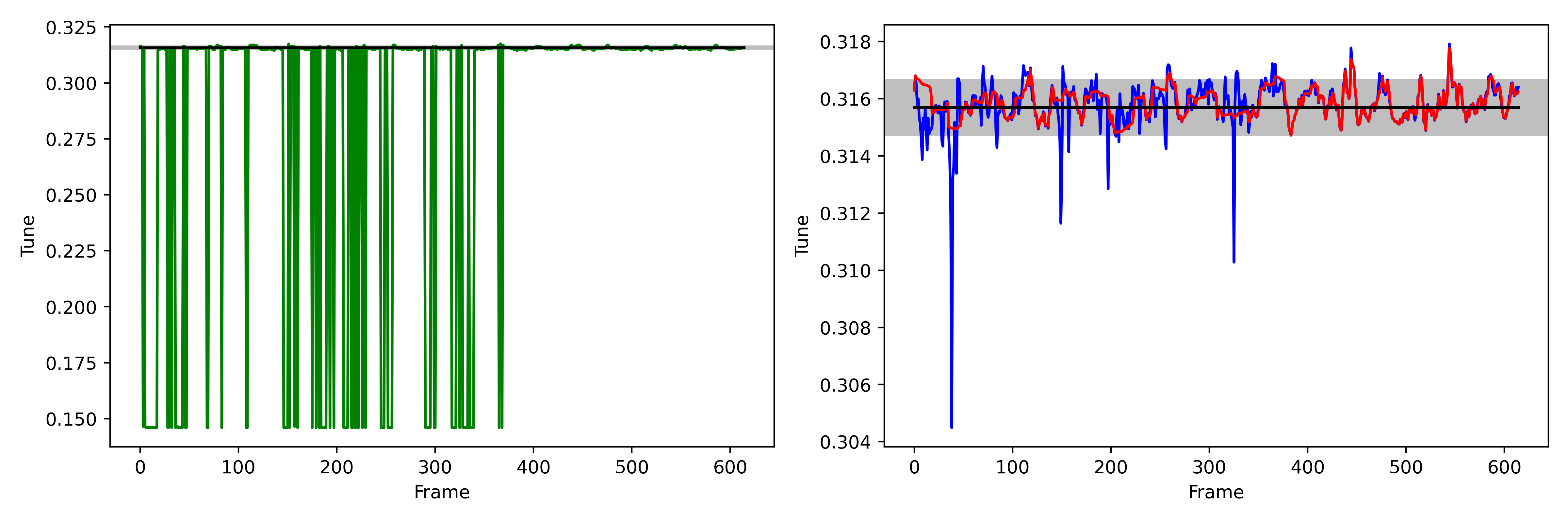}
\caption{Betatron tune tracking on operational beam data during extraction at SAPT (625 frames; reference tune $q_{\text{ref}} = 0.3157$). Left panel: peak-detection baseline. Right panel: neural network without Kalman filtering (blue) versus complete system with uncertainty-aware Kalman filtering (red). The proposed system maintains stable tracking despite occasional large instantaneous errors. Shaded region indicates a $\pm 10^{-3}$ accuracy envelope.}
\label{fig:real_beam_tracking}
\end{figure}

Figure~\ref{fig:real_beam_tracking} shows the temporal evolution of the tune as predicted by three methods: conventional peak detection, the proposed neural network without Kalman filtering, and the complete system combining the neural network with uncertainty-aware Kalman filtering. Both the peak-detection baseline and the raw neural-network outputs exhibit sporadic large deviations from the expected constant tune, with some predictions differing from the reference value by more than 0.01. The uncertainty-aware Kalman filter effectively suppresses these sporadic outliers, producing a smooth tracking trajectory that maintains tune estimates within $1 \times 10^{-3}$ of the reference value throughout the entire sequence. Examination of the predicted uncertainties during outlier events confirms that the network correctly identifies problematic frames: predictions with large instantaneous errors consistently exhibit elevated uncertainty estimates, which trigger automatic down-weighting through the adaptive measurement-noise mechanism.

Quantitatively, the neural network with adaptive Kalman filtering achieves an MAE of $4.48 \times 10^{-4}$ relative to the reference tune over the 625-frame sequence, with 94.96\% of predictions lying within an absolute error of $10^{-3}$. The conventional peak-detection baseline attains 76.59\% compliance with this accuracy tolerance, while the raw neural network achieves an MAE of $5.38 \times 10^{-4}$ with 88.94\% compliance. These results are consistent with the synthetic-validation performance at SNR levels between $-20$ and $-15$~dB, indicating that simulation-based training generalizes effectively to operational beam signals despite domain shifts in noise characteristics and spectral fine structure.

This preliminary validation provides an initial indication of feasibility: neural networks trained entirely on physics-based simulated spectra successfully track betatron tune on real accelerator signals without retraining, fine-tuning, or domain adaptation. However, the limited scope—a single 625-frame sequence during one operational regime at a single facility—precludes comprehensive assessment of generalization across the full range of beam conditions encountered in clinical practice. Future work will require systematic validation across multiple operational scenarios, including injection, acceleration ramps, and varying beam intensities, as well as extended datasets spanning hours to days of operation. Nevertheless, the successful demonstration on authentic beam signals represents a substantial advancement beyond purely simulation-based validation and provides confidence that the proposed diagnostic system is suitable for practical deployment, pending comprehensive commissioning studies.

\section{Conclusions}
\label{sec:conclusions}

This work presents a real-time-capable neural-network-based diagnostic system for betatron tune measurement in medical proton therapy synchrotrons. The system integrates a self-calibrating frequency normalization procedure that eliminates dependence on accurate external revolution-frequency measurements, a lightweight attention-based convolutional neural network that produces calibrated uncertainty estimates, and an uncertainty-aware Kalman filter that exploits frame-specific measurement quality for robust temporal tracking. The complete pipeline achieves sub-millisecond end-to-end latency while providing substantial performance improvements over conventional tune-diagnostic baselines, particularly in challenging low-SNR regimes.

The key contributions are threefold. First, the self-calibrating normalization approach automatically adapts to variations in the revolution frequency, eliminating systematic error from possible desynchronization. Second, the neural network's calibrated uncertainty estimates enable adaptive temporal filtering that substantially outperforms traditional fixed-parameter approaches, with performance gains becoming more pronounced as signal quality degrades. Third, the lightweight architecture maintains real-time performance without specialized hardware acceleration, demonstrating that effective beam diagnostics do not require large-scale models or extensive computational resources.

Comprehensive validation on synthetic data spanning SNR levels from 0 to $-20$~dB demonstrates robust performance across operational conditions. At typical operating SNR ($-15$ to $-10$~dB), the system achieves MAE below $3 \times 10^{-4}$, with uncertainty-aware Kalman filtering providing approximately 20\% error reduction. At the most challenging SNR of $-20$~dB, adaptive filtering achieves approximately 70\% error reduction compared to unfiltered predictions. Preliminary validation on operational beam data from the SAPT facility confirms successful tracking during extraction without retraining.

Future work should prioritize extended experimental validation across diverse operational scenarios, including injection, acceleration ramps, and varying beam intensities. Integration into complete closed-loop control systems and validation of feedback performance under realistic operational conditions are necessary next steps toward clinical deployment. The incorporation of domain-adaptation techniques could enable automatic adjustment to facility-specific characteristics without extensive retraining, while deployment on embedded hardware platforms would demonstrate feasibility for cost-constrained facilities.

\bibliographystyle{JHEP}
\bibliography{main}

\end{document}